\documentclass[showpacs,preprintnumbers,amsmath,amssymb]{revtex4}

\usepackage{graphicx}
\usepackage{dcolumn}
\usepackage{bm}

\begin{document}


\title{PT-Rotations, PT-Spherical Harmonics and the PT-Hydrogen Atom}

\author{Juan M. Romero}
\email{jromero@correo.cua.uam.mx}
\author{R. Bernal-Jaquez}
 \email{rbernal@correo.cua.uam.mx}
 \author{O. Gonz\'alez-Gaxiola}%
 \email{ogonzalez@correo.cua.uam.mx}
\affiliation{%
Departamento de Matem\'aticas Aplicadas y Sistemas\\
Universidad Aut\'onoma Metropolitana-Cuajimalpa, M\'exico 01120 DF,
M\'exico}%

\date{\today}

\begin{abstract}
We have constructed a set of non-Hermitian operators that satisfy the commutation relations of the $SO(3)$-Lie algebra. It is shown that this operators 
generate rotations in the configuration space and not in the momentum space but in a modified non-Hermitian momentum space. 
This generators are related  with a new type of spherical harmonics that result to be $\mathcal{PT}$-orthonormal. Additionally, we have shown that this operators represent conserved quantities for a non-Hermitian Hamiltonian with an additional complex term.  As a particular case, the solutions of the corresponding $\mathcal{PT}$-Hydrogen atom that includes  a complex term are obtained, and it is found that a non-Hermitian Runge-Lenz vector is a conserved quantity. In this way, we obtain a set of non-Hermitian operators that satisfy the  $SO(4)$-Lie  algebra. 
\end{abstract}

\maketitle

\section{Introduction}

Quantum mechanics is considered as one the most solid and well established theories in  physics. Different experiments have corroborated its predictions.  However, at a theoretical level there are different facts that make us think that it could be necessary to modify or extend this theory. For example, it has not been possible to find a consistent quantum-mechanical formulation of general relativity, then quantum theory maybe modified in order to make it compatible with general relativity.\\

As it is well known, quantum mechanics has been formulated in terms of Hermitian operators in order to obtain real spectra. However, it has become clear that Hermiticity is not a necessary condition to obtain real spectra. 
This opens the possibility for quantum mechanics to be extended using non-Hermitian operators, this is the so called $\mathcal{PT}$-symmetry theory, see review \cite{bender-0:gnus}  and its references. \\

The $\mathcal{PT}$-version of quantum mechanics has strongly attracted attention  because it gives a way to deal with some problems that are out of the scope of conventional quantum mechanics. For example, we can solve certain kind of problems in which the  potentials are given by complex-valued functions and whose spectra results to be real \cite{Ben-1:gnus}. In the same way, using this formulation it has been possible to achieve a consistent quantization of a system with high order derivatives: the so called the Pais-Uhlenbeck  oscillator model \cite{bender-uhlenbeck:gnus}. This opens the possibility to construct in a consistent way  high order derivatives  field theories. This fact is important because different theories with high order derivatives have been recently proposed, for example, in extensions of the standard model \cite{pospelov:gnus}, in the noncommutative spaces  \cite{szabo:gnus} and gravity theories  \cite{gravedad:gnus}. In this way,  it becomes possible that an extension of  $\mathcal{PT}$-symmetry theory applied to field theory can give a consistent description of these systems.\\

There is not a finished version of the theory, however, a growing number of themes are under study in the $\mathcal{PT}$-framework, some of them can be found  \cite{das:gnus}. An aspect that has been scarcely treated in the $\mathcal{PT}$-context is the study of symmetries and conserved quantities. In this work, we will study some aspects of this topic. We will obtain a non-Hermitian set of operators that satisfy the commutation relations  of the Lie $SO(3)$ rotation group. It will be shown that these operators generate rotations in the configuration space  $x_i$, and not in the momentum space $\vec{p}=-i \nabla$ but in a modified non-Hermitian momentum space $\vec{p}_f =\vec{p}+i \vec{\nabla{f}}$, originally considered by Dirac in his seminal book \cite{Dirac:gnus}. Also, we will show that the Casimir of the algebra has real spectra and that its eigenfunctions, under the $\mathcal{PT}$-inner product, form a complete basis. This eigenfunctions will be called $\mathcal{PT}$-spherical harmonics. \\

Additionally we will study a central potential Hamiltonian with an additional complex term. It will be shown that the conventional angular momentum is not a conserved quantity anymore and we will have a modified non-Hermitian angular momentum operator. As a particular case, we obtain the solutions of the corresponding $\mathcal{PT}$-Hydrogen atom that includes  a complex term, and it will be found that a non-Hermitian Runge-Lenz vector is a conserved quantity. Then we will have the non-Hermitian generators  of the $SO(4)$-Lie  algebra.\\

This work is organized as follows:  section II  make a brief review of $\mathcal{PT}$-theory and of conventional spherical harmonics, in section III we study the $\mathcal{PT}$-rotations, in section IV  we study the completeness relation and some examples, section V is devoted to the study of symmetry transformations, in section VI  we deal with the central potential problem, in section VII we will study the Hydrogen atom and at last we conclude with a summary of our results.\\\

\section{$\mathcal{PT}$-Symmetry and Spherical Harmonics}

In this section,  a review of some well known facts of  $\mathcal{PT}$-symmetry theory and spherical harmonics is made before consider the $\mathcal{PT}$-transformed version of this functions

\subsection{$\mathcal{PT}$-Inner product}

$\mathcal{PT}$-theory considers the transformations under the parity  operator  $\mathcal{P}$ and the time reversal operator $\mathcal{T}$. Under the $\mathcal{P}$-operator we have the transformation
\begin{eqnarray}
x,y,z\to -x,-y,-z
\end{eqnarray}
and under $\mathcal{T}$
\begin{eqnarray}
i\to -i
\end{eqnarray}
In this way, any function  $f(\vec x)$ can be transform as

\begin{eqnarray}
\mathcal{PT}\left(f(\vec x)\right)=f^{*}(-\vec x).
\end{eqnarray}
Note that, in spherical coordinates  $\mathcal{P}$ produces the transformation
\begin{eqnarray}
(r,\theta,\varphi)\to (r,\pi-\theta,\varphi+\pi),
\end{eqnarray}
under $\mathcal{P}$ a $f$ function transforms as
\begin{eqnarray}
\mathcal{P}\left(f(r,\theta,\varphi)\right) =f(r,\pi-\theta,\varphi+\pi),
\end{eqnarray}
therefore 
\begin{eqnarray}
\mathcal{PT}\left(f(r,\theta,\varphi)\right) =f^{*}(r,\pi-\theta,\varphi+\pi).
\end{eqnarray}
Now, the $\mathcal{PT}$-inner product is defined as  
\begin{eqnarray}
\langle f \vert g\rangle=\int d\vec x[\mathcal{PT}f(x)]g(x).
\end{eqnarray}
This expressions will be used in sections bellow. 
 An exhaustive study of the $\mathcal{PT}$-theory can be found in  \cite{bender-0:gnus}.

\subsection{Spherical Harmonics}

The angular momentum components are given by the Hermitian operators \cite{landau:gnus}
\begin{eqnarray*}
 L_{x}=-i\left( y \frac{\partial }{\partial z}-z \frac{\partial }{\partial y}\right),\;
 L_{y}=-i \left( z \frac{\partial }{\partial x}-x \frac{\partial }{\partial z}\right),\;
L_{z}=-i\left( x \frac{\partial }{\partial y}-y \frac{\partial }{\partial x}\right).
\label{eq:vec-momento-angular}
\end{eqnarray*}
Its algebra is given by
\begin{eqnarray}
\left[L_{x},L_{y}\right]=iL_{z},\qquad \left[L_{z},L_{x}\right]=iL_{y},\qquad 
\left[L_{y},L_{z}\right]=iL_{x}.\label{eq:algebra-lie-0}
\end{eqnarray}
Considering  $L^{2}=L_{x}^{2}+L_{y}^{2}+L_{z}^{2}$ and  Eq. (\ref{eq:algebra-lie-0}), we have
\begin{eqnarray}
\left[L^{2},L_{x}\right]=\left[L^{2},L_{y}\right]=\left[L^{2},L_{z}\right]=0.
\end{eqnarray}
An important equation in mathematical-physics  is the eigenvalue equation $L^{2}Y_{lm}=l(l+1)Y_{lm},l=0,1,2\cdots, $ 
that in spherical coordinates is written
\begin{eqnarray}
L^{2}Y_{lm}(\theta,\varphi)&=& -\left[\frac{1}{\sin\theta} \frac{\partial }{\partial \theta}
\left(\sin\theta\frac{\partial Y_{lm}(\theta,\varphi)  }{\partial \theta}\right)+
\frac{1}{\sin\theta^{2}}\frac{\partial^{2}Y_{lm}(\theta,\varphi) }{\partial \varphi^{2}}\right] \nonumber \\ 
&=& l(l+1)Y_{lm}(\theta,\varphi).
\end{eqnarray}
If $\varphi\in (0,2\pi)$ and  $\theta \in (0,\pi),$ the solutions of this equation are given by the spherical harmonics \cite{fesbach:gnus}
\begin{eqnarray}
Y_{lm}(\theta,\varphi)&=&\sqrt{\frac{(2l+1)(l-m)!}{4\pi (l+m)!}}e^{im\varphi}P_{l}^{m}(\cos \theta), \qquad -l\leq m\leq l,
\label{eq:armonicos-final}
\end{eqnarray}
with
\begin{eqnarray}
l=0, 1, 2,3 \cdots,\qquad 
P_{l}^{m}(u)&=& (-)^{m} (1-u^{2})^{\frac{m}{2}} \frac{d^{m}}{du^{m}}P_{l}(u), \qquad P_{l}(u)=\frac{1}{2^{l}l!}\frac{d^{l}}{du^{l}}\left(u^{2}-1\right)^{l}
\end{eqnarray}
where $P_{l}^{m}(u)$ denoted the associated Legendre polynomials and $P_{l}(u)$ the Legendre polynomials respectively.
The spherical harmonics satisfy the orthonormality relation 
\begin{eqnarray}
<Y_{l^{\prime}m^{\prime}}(\theta,\varphi)|Y_{l m}(\theta,\varphi)>
=\int d\Omega Y^{*}_{l^{\prime}m^{\prime}}(\theta,\varphi) 
Y_{l m}(\theta,\varphi)=\delta_{mm^{\prime}}\delta_{l^{\prime}l}.\label{eq:p80}
\end{eqnarray}
Given that the spherical harmonics constitute an orthonormal basis,  we can write any function $F(\theta,\varphi)$ as a linear combination of them, that is
\begin{eqnarray}
F(\theta, \varphi)=\sum_{l\geq 0} \sum_{m=-l}^{l}C_{lm}
Y_{lm}(\theta, \varphi).
\label{eq:serie-arm}
\end{eqnarray}
Using  Eq. (\ref{eq:p80}), we find
\begin{eqnarray}
C_{lm}=\int d\Omega Y^{*}_{lm}(\theta, \varphi) F(\theta, \varphi).
\end{eqnarray}
Substituting  $C_{lm}$ in Eq. (\ref{eq:serie-arm}) and making the change of variables $u^{\prime}=\cos\theta^{\prime},u=\cos\theta$ 
we obtain
\begin{eqnarray}
F(\theta, \varphi)=\int_{0}^{2\pi} d\varphi\int_{-1}^{1}du^{\prime} 
F(u^{\prime}, \varphi^{\prime}) 
\left(\sum_{l\geq 0} \sum_{m=-l}^{l} 
Y^{*}_{lm}(u^{\prime}, \varphi^{\prime})Y_{lm}(u, \varphi)\right).\nonumber
\end{eqnarray}
Therefore, the expression inside the parenthesis must be equal to
$\delta(\varphi -\varphi^{\prime})\delta(u -u^{\prime}),$ that is
\begin{eqnarray}
\sum_{l\geq 0}\sum_{m=-l}^{l}Y^{*}_{lm}(\theta^{\prime},\phi^{\prime})
Y_{lm}(\theta,\phi)=\delta(\phi-\phi^{\prime}) 
\delta(\cos\theta-\cos\theta^{\prime} ), 
\label{eq:completez-arm}
\end{eqnarray}
this expression is called completeness relation.\\

Under the parity operator the spherical harmonics transform as
\begin{eqnarray}
\mathcal{P}\left(Y_{l m}(\theta,\varphi)\right)=Y_{l m}(\pi-\theta,\varphi+\pi)=(-)^{l}Y_{l m}(\theta,\varphi),
\end{eqnarray}
and under the $\mathcal{PT}$-operator, we have
\begin{eqnarray}
\mathcal{PT}\left(Y_{l m}(\theta,\varphi)\right)=Y^{*}_{l m}(\pi-\theta,\varphi+\pi)=(-)^{l}Y^{*}_{l m}(\theta,\varphi).
\end{eqnarray}
This results will be used below.

\section{$\mathcal{PT}$ and Rotations}

Given any $f=f(r,\theta,\varphi)$,  we can define
\begin{eqnarray}
L_{fi}=e^{f}L_{i}e^{-f}.
\end{eqnarray}
In general $L_{fi}$ is a non-Hermitian operator, however
\begin{eqnarray}
\left[L_{fx},L_{fy}\right]=iL_{fz},\qquad \left[L_{fz},L_{fx}\right]=iL_{fy},\qquad 
\left[L_{fy},L_{fz}\right]=iL_{fx},\label{eq:algebra-lie}
\end{eqnarray}
that we can identify as the $SO(3)$-Lie algebra commutation relations.
Considering  $L_{f}^{2}=L_{fx}^{2}+L_{fy}^{2}+L_{fz}^{2}$ and Eq. (\ref{eq:algebra-lie}),
we have
\begin{eqnarray}
\left[L_{f}^{2},L_{fi}\right]=0,
\end{eqnarray}
therefore, we have the same algebra as the one satisfied by  $L_{i}.$ Besides
\begin{eqnarray}
L_{f}^{2}Y_{flm}(\theta,\varphi)=l(l+1)Y_{flm}(\theta,\varphi),\qquad Y_{flm}(\theta,\varphi)=e^{f}Y_{lm}(\theta,\varphi),
\end{eqnarray}
that will be called $\mathcal{PT}$-spherical harmonics.
In this case, the $\mathcal{PT}$-inner product is given by
\begin{eqnarray}
<Y_{fl^{\prime}m^{\prime}}(\theta,\varphi)|Y_{fl m}(\theta,\varphi)>_{f}
&=&\int d\Omega \mathcal{PT}\left(Y_{fl^{\prime}m^{\prime}}(\theta,\varphi)\right) Y_{fl m}(\theta,\varphi).
\end{eqnarray}
Under a $\mathcal{PT}$-transformation, we have 
\begin{eqnarray}
\mathcal{PT}(Y_{flm})&=& e^{f^{*}(r,\pi-\theta,\varphi+\pi)}(-)^{l}Y^{*}_{lm}(\theta,\varphi).
\end{eqnarray}
therefore we can write

\begin{eqnarray}
<Y_{fl^{\prime}m^{\prime}}(\theta,\varphi)|Y_{fl m}(\theta,\varphi)>_{f}
&=& (-)^{l} \int d\Omega e^{f^{*}(r,\pi-\theta,\varphi+\pi)+f(r,\theta,\varphi)}Y^{*}_{l^{\prime}m^{\prime}}(\theta,\varphi)Y_{lm}(\theta,\varphi).
\nonumber
\end{eqnarray}
It becomes clear that, under this inner product not any function $f$ allows  the set $Y_{flm}$ to be an orthogonal set. However, if the following condition is fulfilled
\begin{eqnarray}
e^{f^{*}(r,\pi-\theta,\varphi+\pi)+f(r,\theta,\varphi)}=\lambda,\qquad \lambda={\rm const}
\label{eq:con-pt}
\end{eqnarray}
then we have 
\begin{eqnarray}
<Y_{fl^{\prime}m^{\prime}}(\theta,\varphi)|Y_{fl m}(\theta,\varphi)>_{f}
&=&(-)^{l}\lambda \delta_{l^{\prime}l}\delta_{m^{\prime}m}.
\label{eq:p80f}
\end{eqnarray}
In this way, the spherical harmonics  $Y_{flm}(\theta,\varphi)$ are orthogonal under the $\mathcal{PT}$-inner product  only  if  Eq. (\ref{eq:con-pt}) is satisfy.
It is worthy to mention that due to the parity of  the wave functions, in some $\mathcal{PT}$-symmetry systems  the following orthogonality relations 
\begin{eqnarray}
\langle \phi_{m}|\phi_{n}\rangle=(-1)^{n}\delta_{m,n},
\end{eqnarray}
may be obtained \cite{bender-0:gnus}.
\section{Completeness Relation}

Using the $\mathcal{PT}$-spherical harmonics $Y_{flm}(\theta,\varphi)$, we can have the expansion 
\begin{eqnarray}
F(\theta, \varphi)=\sum_{l\geq 0} \sum_{m=-l}^{l}a_{lm}
Y_{flm}(\theta, \varphi).
\label{eq:serie-arm-f}
\end{eqnarray}
Appeling to the orthonormality relations Eq. (\ref{eq:p80f}), we find
\begin{eqnarray}
a_{lm}=\frac{(-)^{l}}{\lambda} <Y_{flm}(\theta,\varphi)|F(\theta,\varphi)>_{f}=\frac{(-)^{l}}{\lambda}\int d\Omega \mathcal{PT}\left(Y_{flm}(\theta, \varphi)\right) F(\theta, \varphi).
\end{eqnarray}
substituting this result into  Eq.(\ref{eq:serie-arm-f}), we obtain 
\begin{eqnarray}
F(\theta, \varphi)&=&\sum_{l\geq 0} \sum_{m=-l}^{l}\frac{(-)^{l}}{\lambda}\int d\Omega^{\prime} \mathcal{PT}\left(Y_{flm}(\theta^{\prime}, \varphi^{\prime})\right) F(\theta^{\prime}, \varphi^{\prime})
Y_{flm}(\theta, \varphi)\nonumber\\
 &=& \int d\Omega^{\prime} F(\theta^{\prime}, \varphi^{\prime})\left[ 
 \sum_{l\geq 0} \sum_{m=-l}^{l}\frac{(-)^{l}}{\lambda} \mathcal{PT}\left(Y_{flm}(\theta^{\prime}, \varphi^{\prime})\right)Y_{flm}(\theta, \varphi) \right]\nonumber\\
&=& \int d\Omega^{\prime} F(\theta^{\prime}, \varphi^{\prime})\left[ 
 \sum_{l\geq 0} \sum_{m=-l}^{l}\frac{ e^{f^{*}(r,\pi-\theta^{\prime},\varphi^{\prime}+\pi)+f(r,\theta,\varphi)}} {\lambda} Y^{*}_{lm}(\theta^{\prime}, \varphi^{\prime})Y_{lm}(\theta, \varphi) \right],\nonumber
\end{eqnarray}
therefore
\begin{eqnarray}
 \delta(\phi-\phi^{\prime}) 
\delta(\cos\theta-\cos\theta^{\prime} )=
\sum_{l\geq 0} \sum_{m=-l}^{l}\frac{ e^{f^{*}(r,\pi-\theta^{\prime},\varphi^{\prime}+\pi)+
f(r,\theta,\varphi)}} {\lambda} Y^{*}_{lm}(\theta^{\prime}, \varphi^{\prime})Y_{lm}(\theta, \varphi).
\end{eqnarray}
This is the completeness relation for the $\mathcal{PT}$-spherical harmonics. A similar completeness relation is found for different systems in $\mathcal{PT}$-quantum mechanics \cite{bender-0:gnus}.

\subsection{Examples on Completeness}

In this subsection we will see some examples of functions that satisfy Eq. (\ref{eq:con-pt}).\\

Let us suppose that $a$ is a real constant, then we can define
\begin{eqnarray}
f(r,\theta,\varphi)=a\theta
\end{eqnarray}
therefore
\begin{eqnarray}
e^{f^{*}(r,\pi-\theta,\varphi+\pi)+f(r,\theta,\varphi)}=e^{a(\pi-\theta)+a\theta}=e^{a\pi},
\end{eqnarray}
where
\begin{eqnarray}
\lambda =e^{a\pi}.
\end{eqnarray}
In this case, the orthonormality relations are given by
\begin{eqnarray}
<Y_{fl^{\prime}m^{\prime}}(\theta,\varphi)|Y_{fl m}(\theta,\varphi)>_{f}
&=&(-)^{l}e^{a\pi}\delta_{l^{\prime}l}\delta_{m^{\prime}m}
\end{eqnarray}
and the completeness relation is given by
\begin{eqnarray}
 \delta(\phi-\phi^{\prime}) 
\delta(\cos\theta-\cos\theta^{\prime} )=
\sum_{l\geq 0} \sum_{m=-l}^{l}e^{a(\theta-\theta^{\prime})} Y^{*}_{lm}(\theta^{\prime}, \varphi^{\prime})Y_{lm}(\theta, \varphi).
\end{eqnarray}

Now consider
\begin{eqnarray}
f(r,\theta,\varphi)=ai\sin\theta,
\end{eqnarray}
then we obtain
\begin{eqnarray}
e^{f^{*}(r,\pi-\theta,\varphi+\pi)+f(r,\theta,\varphi)}=e^{-ai\sin\theta+ia\sin\theta}=1=\lambda
\end{eqnarray}
with the orthonormality relations
\begin{eqnarray}
<Y_{fl^{\prime}m^{\prime}}(\theta,\varphi)|Y_{fl m}(\theta,\varphi)>_{f}
&=&(-)^{l} \delta_{l^{\prime}l}\delta_{m^{\prime}m}.
\end{eqnarray}
In this case the completeness relation is given by
\begin{eqnarray}
 \delta(\phi-\phi^{\prime}) 
\delta(\cos\theta-\cos\theta^{\prime} )=
\sum_{l\geq 0} \sum_{m=-l}^{l}e^{ai(\sin\theta-\sin\theta^{\prime})} Y^{*}_{lm}(\theta^{\prime}, \varphi^{\prime})Y_{lm}(\theta, \varphi).
\end{eqnarray}
\\
Using
\begin{eqnarray}
f(r,\theta,\varphi)=a\cos\theta,
\end{eqnarray}
we have
\begin{eqnarray}
e^{f^{*}(r,\pi-\theta,\varphi+\pi)+f(r,\theta,\varphi)}=e^{-a\cos\theta+a\cos\theta}=\lambda=1
\end{eqnarray}
in this case the orthogonality relations are given by
\begin{eqnarray}
<Y_{fl^{\prime}m^{\prime}}(\theta,\varphi)|Y_{fl m}(\theta,\varphi)>_{f}
&=&(-)^{l} \delta_{l^{\prime}l}\delta_{m^{\prime}m},
\end{eqnarray}
and the completeness relation is given by
\begin{eqnarray}
 \delta(\phi-\phi^{\prime}) 
\delta(\cos\theta-\cos\theta^{\prime} )=
\sum_{l\geq 0} \sum_{m=-l}^{l}e^{a(\cos\theta-\cos\theta^{\prime})} Y^{*}_{lm}(\theta^{\prime}, \varphi^{\prime})Y_{lm}(\theta, \varphi).
\end{eqnarray}
\\
With the function 
\begin{eqnarray}
f(r,\theta,\varphi)=ai\varphi
\end{eqnarray}
we obtain 
\begin{eqnarray}
e^{f^{*}(r,\pi-\theta,\varphi+\pi)+f(r,\theta,\varphi)}=e^{-ai(\varphi+\pi)+ ia\varphi}=e^{-ia\pi}=\lambda,
\end{eqnarray}
with the orthogonality relation given by
\begin{eqnarray}
<Y_{fl^{\prime}m^{\prime}}(\theta,\varphi)|Y_{fl m}(\theta,\varphi)>_{f}
&=&(-)^{l}e^{-ia\pi} \delta_{l^{\prime}l}\delta_{m^{\prime}m},
\end{eqnarray}
and the completeness relation
\begin{eqnarray}
 \delta(\phi-\phi^{\prime}) 
\delta(\cos\theta-\cos\theta^{\prime} )=
\sum_{l\geq 0} \sum_{m=-l}^{l}e^{ai(\varphi-\varphi^{\prime})} Y^{*}_{lm}(\theta^{\prime}.\varphi^{\prime})Y_{lm}(\theta, \varphi).
\end{eqnarray}
Considering the above examples, it is clear that many functions  satisfy Eq. (\ref{eq:con-pt}).

\section{Symmetry Transformations}

Consider $A,B,C$ that satisfy the following commutation relations
\begin{eqnarray}
[A,B]=C.
\end{eqnarray}
Transforming the $A,B,C$  operators,  we obtain $A_{f}=e^{f}Ae^{-f},B_{f}=e^{f}Be^{-f}$ y $C_{f}=e^{f}Ce^{-f},$
then 
\begin{eqnarray}
[A_{f},B_{f}]=C_{f}.
\end{eqnarray}
We know that $[L_{i},x_{j}]=i\epsilon_{ijk}x_{k}$. If we consider the transformation $x_{fi}=e^{f}x_{i}e^{-f}=x_{i}$, we arrive to 
\begin{eqnarray}
[L_{fi},x_{j}]=i\epsilon_{ijk}x_{k},
\end{eqnarray}
therefore the operators $L_{fi}$ generate infinitesimal rotations in the space $x_{i}$. However,  as $p_{i}=-i\frac{\partial }{\partial x^{i}},$ 
then
\begin{eqnarray}
[L_{fi},p_{j}]\not =i\epsilon_{ijk}p_{k},
\end{eqnarray}
and we say that the operators $L_{fi}$ does not generate infinitesimal rotations in the space $p_{i}$. Now, if $p_{fi}$  is  given by  $p_{fi}=e^{f}p_{i}e^{-f}$, then
we have 
\begin{eqnarray}
[L_{fi},p_{fj}] =i\epsilon_{ijk}p_{fk}.
\end{eqnarray}
Note that  
\begin{eqnarray}
\vec p_{f} = e^{f}\vec p e^{-f}=\vec p+i\vec \nabla f.
\end{eqnarray}
this operator was studied by Dirac in his seminal book \cite{Dirac:gnus}.
If the Hamiltonian operator is given by
\begin{eqnarray}
H=\frac{\vec p^{\;2}}{2m}+V(r), 
\end{eqnarray}
then
\begin{eqnarray}
[L_{i},H] =0.
\end{eqnarray}
Defining $H_f$ by 
\begin{eqnarray}
H_{f} = e^{f} H e^{-f},
\end{eqnarray}
we have that, in general
\begin{eqnarray}
[L_{i},H_{f}] \not=0,
\end{eqnarray}
therefore the angular momentum  $L_{i}$ is not a conserved quantity for Hamiltonians of the form $H_{f}.$
However
\begin{eqnarray}
[L_{fi},H_{f}] =0,
\end{eqnarray}
then the modified angular momentum $L_{fi}$ is conserved.
Note that the Hamiltonian $H_f$  is given by
\begin{eqnarray}
H_{f}=\frac{\vec p_{f}^{\;2}}{2m}+V(r)= \frac{m}{2}\left(\vec p+i \vec \nabla f\right)^{2}+V(r)
\end{eqnarray}
In the next section we will consider one important example.

\section{The Central Problem}

Consider the Hamiltonian 
\begin{eqnarray}
H=\frac{m}{2}\vec{p}^{\;2}+V(x,y,z),
\end{eqnarray}
then
\begin{eqnarray}
H_{f}&=& e^{f}H e^{-f}=\frac{m}{2}\left(\vec p+i \vec \nabla f\right)^{2}+V(x,y,z)\nonumber\\
&=&\frac{m}{2}\left(\vec p^{\;2}+2i\vec \nabla f\cdot \vec p+(\nabla^{2}f)-\left(\vec \nabla f\right)^{2}\right)+ V(x,y,z),
\end{eqnarray}
that is a  non-Hermitian Hamiltonian. If the potential is given by
\begin{eqnarray}
V(x,y,z)=-\frac{m}{2}\left( \nabla^{2}f-\left(\vec \nabla f\right)^{2}\right)+ U(x,y,z),
\end{eqnarray}
then we can write 
\begin{eqnarray}
H_{f}&=& \frac{m}{2}\left(\vec p^{\;2}+2i\vec \nabla f\cdot \vec p\right)+ U(x,y,z).
\end{eqnarray}
This kind of Hamiltonians naturally arise in some statistical models \cite{reichl:gnus}.\\

Note that if $\psi$ is an eigenfunction in the equation
\begin{eqnarray}
H\psi=E\psi
\end{eqnarray}
then we can define the $f$-states  $\psi_{f}=e^{f}\psi$ that satisfy 
\begin{eqnarray}
H_{f}\psi_{f}=E\psi_{f}.
\end{eqnarray}
It follows that although $H_{f}$ is a non-Hermitian operator it does has a real spectrum.\\

As an example, let us consider the central potential problem $V(r)$ whose  Schrodinger equation is given by \cite{landau:gnus} 
\begin{eqnarray}
H\psi=\left( \frac{m}{2}\vec p^{\;2}+V(r)\right)\psi= E\psi
\end{eqnarray}
and whose solutions 
\begin{eqnarray}
\psi_{E}(r,\theta,\varphi)=\phi_{E}(r)Y_{lm}(\theta,\varphi)
\end{eqnarray}
satisfy the orthogonality relations
\begin{eqnarray}
<\psi_{E^{\prime}}(r,\theta,\varphi)|\psi_{E}(r,\theta,\varphi)>=
\int dr r^{2}d\Omega \psi^{*}_{E^{\prime}}(r,\theta,\varphi)\psi_{E}(r,\theta,\varphi)=
\delta_{EE^{\prime}}.\label{eq:orto-radial}
\end{eqnarray}
Then the solutions of the equation  
\begin{eqnarray}
H_{f}\psi_{f}&=& \left[\frac{m}{2}\left(\vec{p}^{\;2}+2i\vec \nabla f\cdot \vec p+ \nabla^{2}f -\left(\vec \nabla f\right)^{2}      \right)+ V(r)\right]\psi_{f} \nonumber \\ 
&=& E\psi_{f} \label{rancherita}
\end{eqnarray}
are given by
\begin{eqnarray}
\psi_{Ef}(r,\theta,\varphi)=e^{f(r,\theta,\varphi)}\phi_{E}(r)Y_{lm}(\theta,\varphi).
\end{eqnarray}
The $\mathcal{PT}$-inner product for the $\psi_f$-states is given by
\begin{eqnarray}
& &<\psi_{E^{\prime}f}(r,\theta,\varphi)|\psi_{Ef}(r,\theta,\varphi)>_{f}=\int dr d\Omega \mathcal{PT}\left( \psi_{E^{\prime}f}(r,\theta,\varphi)\right)\psi_{Ef}(r,\theta,\varphi)\nonumber\\
 &=&\int dr d\Omega (-)^{l}e^{f^{*}(r,\pi-\theta,\varphi+\pi)+f (r,\theta,\varphi) } 
 \phi_{E^{\prime}}^{*}(r)
 \phi_{E}(r) Y^{*}_{l^{\prime} m^{\prime} }(\theta,\varphi) Y_{lm}(\theta,\varphi)\nonumber.
\end{eqnarray}
Besides, if  $f(r,\theta,\varphi)$ satisfies Eq. (\ref{eq:con-pt}) and  considering Eq. (\ref{eq:orto-radial}), we have
\begin{eqnarray}
<\psi_{E^{\prime}f}(r,\theta,\varphi)|\psi_{Ef}(r,\theta,\varphi)>_{f}
&=&(-)^{l}\lambda \int dr d\Omega  \phi_{E^{\prime}}^{*}(r)
 \phi_{E}(r) Y^{*}_{l^{\prime} m^{\prime} }(\theta,\varphi) Y_{lm}(\theta,\varphi)\nonumber\\
&=&\lambda (-)^{l}\delta_{EE^{\prime}}.
\end{eqnarray}
Given that  $[L_{i},H_{f}]\not =0$, it follows that $L_{i}$ is not a conserved quantity. However, as
\begin{eqnarray}
[L_{fi},H_{f}] =0,
\end{eqnarray}
then $L_{fi}$  is conserved.

\section{The Hydrogen Atom}

In the case of the Hydrogen atom, we have the potential
\begin{eqnarray}
V(r)=-\frac{Ze^{2}}{r},
\end{eqnarray}
where the solutions are given by 
\begin{eqnarray}
\psi_{Nlm}(\rho,\theta,\varphi)&=&\frac{2}{N^{2}}\sqrt{\frac{Z^{3}}{a_{RB}^{3}}  \frac{(N-l-1)!}{(N+l)!}}
\rho^{l} L_{N-(l+1)}^{2l+1}(\rho)e^{-\frac{\rho}{2}}Y_{lm}(\theta,\varphi),\nonumber\\
E_{N}&=& -\frac{Ze^{2}}{a_{RB}N^{2}},\qquad N=n+l+1,\qquad n=0,1,2\cdots ,\nonumber
\end{eqnarray}
where $a_{RB}$ is the Bohr radius and  
\begin{eqnarray}
\rho=\alpha r,\qquad 
\qquad \alpha=2 \sqrt{\frac{-2mE}{\hbar^{2}} }.
\end{eqnarray}
Taking into account Eq. (\ref{rancherita}), we have the equation 
\begin{eqnarray}
H_{f}\psi_{f}&=& \left[\frac{m}{2}\left(\vec{p}^{\;2}+2i\vec \nabla f\cdot \vec{p}+ \nabla^{2}f  -\left(\vec \nabla f\right)^{2} \right)-\frac{Ze^{2}}{r}\right]\psi_{f}=E\psi_{f}
\end{eqnarray}
whose solutions are given by 
\begin{eqnarray}
\psi_{fNlm}(\rho,\theta,\varphi)&=& e^{f(r,\theta,\varphi)} \psi_{Nlm}(\rho,\theta,\varphi)
\end{eqnarray}
this are  orthogonal  functions if  equation (\ref{eq:con-pt}) is satisfied. A remarkable fact is that  $L_{fi}$ is a conserved quantity.
In the conventional Hydrogen atom, the Runge-Lenz vector is also conserved  \cite{landau:gnus}
\begin{eqnarray}
R_{i}= \frac{1}{2}\left(\vec L\times \vec p-\vec p\times \vec L+\frac{Ze^{2}\vec r}{r}\right)_{i}.
\end{eqnarray}
In the case of the Hamiltonian  $H_{f}$, we have
\begin{eqnarray}
[R_{fi},H_{f}] =0,
\end{eqnarray}
and we can say that the transformed non-Hermitian Runge-Lenz vector is conserved.\\
Note that in this case,  we have obtained a set of conserved quantities $L_{fi}, L_{f}^2, R_{fi}$, that are the non-Hermitian generators of the $SO(4)$ algebra. \\

\section{Summary}
In this work we have constructed a set of non-Hermitian operators $L_{fi}$ that satisfy the commutation relations of the $SO(3)$-Lie algebra. We have shown that this operators generate rotations in the configuration space
and not in the conventional momentum space but in a modified non-Hermitian momentum space $\vec{p}_f =\vec{p}+i \vec{\nabla{f}}.$ It is worthy to mention that this operator was originally considered by Dirac in his seminal book.
Besides, the $L_{if}$ generators are related  with a new type of spherical harmonics that result to be $\mathcal{PT}$-orthonormal. Additionally, we have shown that this quantities are conserved for mechanical systems described by  a central potential Hamiltonian with an additional complex term.  As a particular case, we have obtained  the solutions of the corresponding $\mathcal{PT}$-Hydrogen atom that includes  a complex term, and we have found that a non-Hermitian Runge-Lenz vector is a conserved quantity. Considering this case, one remarkable result is that, as we have obtained the non-Hermitian generators of the $SO(3)$-Lie algebra and also a non-Hermitian Runge-Lenz vector,  then we have the non-Hermitian generators  of the $SO(4)$-Lie  algebra. 
In a future work  we will study the non-Hermitian generators corresponding to others symmetry groups.

\end{document}